\documentclass[aps,prl,reprint,superscriptaddress]{revtex4-1}

\usepackage{graphicx}
\usepackage{url}
\usepackage{bm}
\usepackage{color}
\usepackage{textcomp}
\usepackage{lmodern}
\usepackage{xfrac}
\usepackage{longtable}
\usepackage{float}

\begin{document}

\title{Stable Polar Oxynitrides through Epitaxial Strain}


\author{Li Zhu}
\email{z@zhuli.name}
\affiliation{Extreme Materials Initiative, Earth and Planets Laboratory, Carnegie Institution for Science, 5241 Broad Branch Road, NW, Washington, DC 20015, USA}
\altaffiliation{Current address: Department of Physics, Rutgers University, Newark, NJ 07102, USA}

\author{Hiroyuki Takenaka}
\affiliation{Extreme Materials Initiative, Earth and Planets Laboratory, Carnegie Institution for Science, 5241 Broad Branch Road, NW, Washington, DC 20015, USA}
\affiliation{Department of Chemistry and Biochemistry
University of California, Santa Cruz, CA 95060, USA}

\author{R. E. Cohen}
\email{rcohen@carnegiescience.edu}
\affiliation{Extreme Materials Initiative, Earth and Planets Laboratory, Carnegie Institution for Science, 5241 Broad Branch Road, NW, Washington, DC 20015, USA}

\date{\today}

\begin{abstract}

We investigate energetically favorable structures of ABO$_2$N oxynitrides as functions of pressure and strain via swarm-intelligence-based structure prediction methods, DFT lattice dynamics and first-principles molecular dynamics. We predict several thermodynamically stable polar oxynitride perovskites under high pressures. In addition, we find that ferroelectric polar phases of perovskite-structured oxynitrides can be thermodynamically stable and synthesized at high pressure on appropriate substrates. The dynamical stability of the ferroelectric oxynitrides under epitaxial strain at ambient pressure also imply the possibility to synthesize them using pulsed laser deposition or other atomic layer deposition methods. Our results have broad implications for further exploration of other oxynitride materials as well. We performed first-principles molecular dynamics and find that the polar perovskite of YSiO$_2$N is metastable up to at least 600 K under compressive epitaxial strain before converting to the stable wollastonite-like structures. YSiO$_2$N is stabilized under pressure with extensional epitaxial strain. We predict that LaSi$_2$N, LaGeO$_2$N, BiSiO$_2$N, and BiGeO$_2$N are metastable as ferroelectric perovskites at zero pressure even without epitaxial strain.
\end{abstract}


\maketitle


The perovskite oxides (ABO$_3$) form one of the most widely studied groups in condensed matter physics and materials science. Extensive studies over decades show that perovskite oxides possess an exceptional diversity of physical and chemical properties~\cite{Okuda2001,Saghi-Szabo1998,Kutnjak2006,Wu1987}.
Perovskite oxides are particularly important as ferroelectrics in numerous applications ranging from medical ultrasound to sonar~\cite{Haertling1999,Wessels2007,Scott2007}.
The large diversity of perovskites oxides could be further increased by anion substitution. For example, the photovoltaic performance can be boosted through halide anions substitution for oxygen in oxide perovskites~\cite{Jena2019}.
As another example, nitrogen substitution for oxygen enriches the possible perovskite structures and their properties due to the concomitant interaction between oxygen and nitrogen ions~\cite{Fuertes2012}. The investigation of perovskite oxynitrides (ABO$_2$N) has rapidly become a highly important subject area because they represent an emerging class of materials offering the prospect of optimized properties and potential applications in many field, such as visible-light photocatalysts for water splitting, non-toxic pigments, and colossal magnetoresistance~\cite{Kim2004,Withers2008,Hinuma2012,Yajima2015,LePaven-Thivet2009,Higashi2013,Balaz2013a,Jansen2000}. The different ionicities/covalencies between O and N ions may also induce the formation of strong polar perovskite structures, which hold the potential for piezoelectric and ferroelectric applications. In 2007, Caracas and Cohen predicted the polar ordered oxynitride perovskite YSiO$_2$N with high predicted spontaneous polarization and large non-linear optic coefficient~\cite{Caracas2007}.
This phase was synthesized in a diamond anvil cell from YN and SiO$_2$ in 2017~\cite{Vadapoo2017}.

Much oxynitride synthesis is done by ammoniazation of oxides, which results in random substitution of O by N, resulting in non-polar structure~\cite{Morgan1977,Jansen2000,Kim2004}. Finding suitable starting materials for solid state synthesis is a challenge, as N-bonds are either very strong or very unstable. For decades, pressure has been widely used as a powerful tool in the discovery of materials inaccessible at ambient conditions. Thus YSiO$_2$N was synthesized in a polar perovskite phase under high temperature and high pressure conditions~\cite{Vadapoo2017}.

Despite the possibility for many novel phases with enhanced functionality, there are very limited studies on the exploration of pressure-induced polar structures in oxynitrides, leaving the rationally design of the synthesizable polar oxynitride perovskites unsettled. To tackle this problem, it is crucial to understand the thermodynamically stable structures and their properties, which has become possible becasue of the development of crystal structure prediction methods. In this work, we first predicted the high-pressure phase diagrams of various oxynitride systems using swarm-intelligence-based structure prediction methods~\cite{Wang2010,Wang2012a,GAO2019301}.
The predictions reveal several pressure-induced polar oxynitrides. Further calculations indicate that the long-thought ferroelectric phase of oxynitrides can be stable and synthesized under epitaxial strain conditions.

\begin{figure*}
\includegraphics[width=17cm]{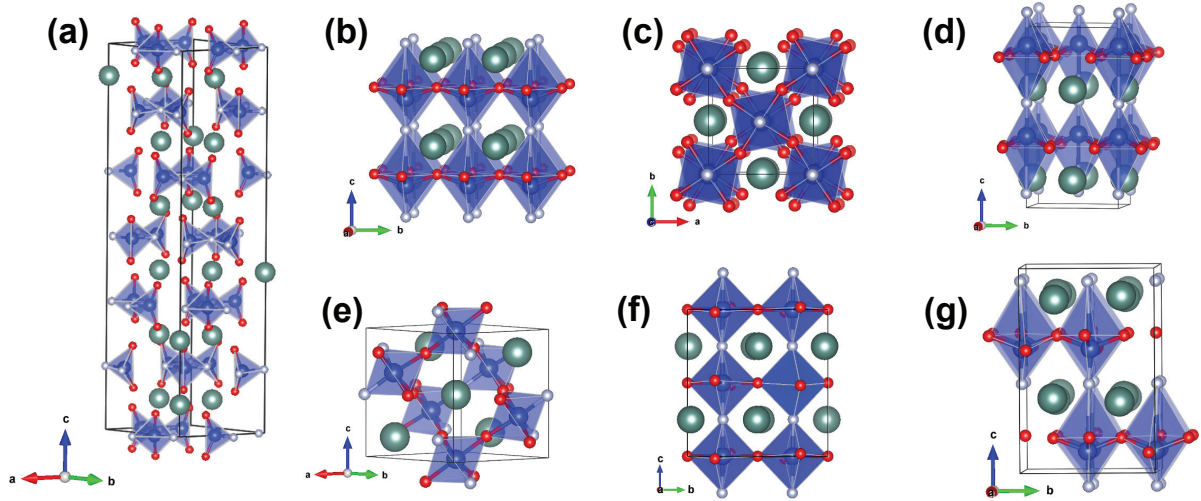}
\caption{\label{fig1} Crystal structures of oxynitride compounds in (a) the $P6_122$ structure, (b) the $P4mm$ structure, (c) the $I4/mcm$ structure, (d) the $I4cm$ structure, (e) the $P3_121$ structure, (f) the $P3_2$ structure, and (g) the $Cmc2_1$ structure.  Cyan, blue, red, and light grey spheres represent Y/La/Bi, Si/Ge/Hf, O, and Y atoms, respectively. } 
\end{figure*}

\begin{figure*}
    \includegraphics[width=17cm]{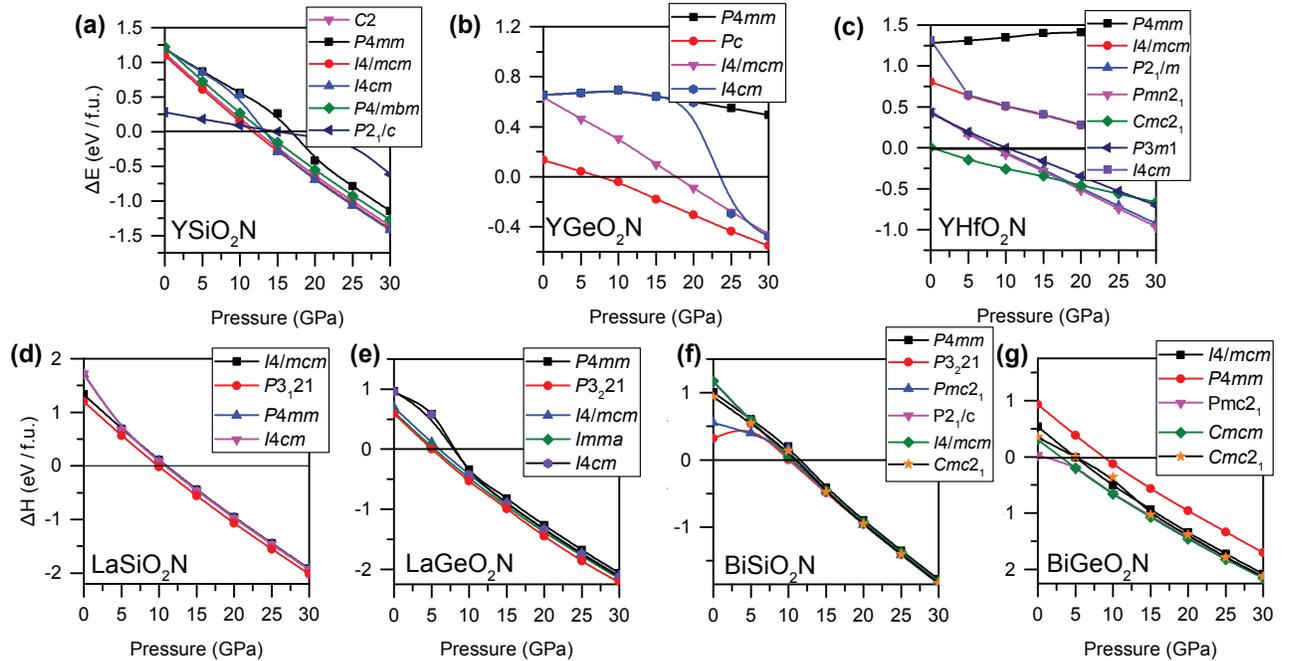}
        \caption{\label{fig2} Calculated enthalpies per formula unit (f.u.) of various predicted structures as functions of pressure with respect to the $P6_122$ structure of the oxynitrides.} 
\end{figure*}

Global structural optimization was performed using the \textsc{calypso} code~\cite{Wang2010,Wang2012a,GAO2019301} with the particle swarm optimization algorithm, which has successfully predicted structures of various systems~\cite{Zhu2011,Zhu2012, Zhu2014,Wang2015,Zhu2019a,Zhu2020}. The energetic calculations were carried out in the framework of density functional theory (DFT) within the Perdew-Burke-Ernzerhof revised for solids (PBEsol)~\cite{Perdew2008} functional, as implemented in the Vienna ab initio Simulation Package (\textsc{vasp}) code~\cite{Kresse1996}. The previous study verified that the PBEsol functional can reproduce the experimental results very well compared to the other functionals~\cite{PhysRevB.89.064305}. The all-electron projector augmented wave (PAW) method~\cite{Blochl1994} as adopted with the pseudopotentials taken from the VASP library where $4s^24p^64d^15s^2$, $5s^25p^65d^16s^2$, $5d^{10}6s^26p^3$, $3s^23p^2$, $4s^24p^2$, $5d^26s^2$, $2s^22p^4$, and $2s^22p^5$ were treated as valence electrons for Y, La, Bi, Si, Ge, Hf, O, and N, respectively. The electronic wave functions were expanded in a plane-wave basis set with a cutoff energy of 520 eV. Monkhorst-Pack $k$-point meshes~\cite{Monkhorst1976} with a grid of spacing 0.04 $\times$ 2$\pi$ {\AA}$^{-1}$ for Brillouin zone sampling were chosen after checking for convergence. For example, we used a $8\times8\times6$ $k$-point for the 5-atom perovskite structures. To determine the dynamical stability of the studied structures, we performed phonon calculations by using the finite displacement approach, as implemented in the \textsc{phonopy} code~\cite{Togo2015a}.

Structure prediction calculations were performed at a pressure of 30 GPa with up to four formula units (f.u.) per simulation cell. We uncovered a group of new structures for the oxynitrides at 30 GPa. We take the crystal phases identified by the structure search process and compute their enthalpies to determine the most stable structure for each composition. The enthalpies were calculated by optimizing the cell parameters and the atomic positions at each pressure using the conjugate gradient algorithm implemented in the VASP code~\cite{Kresse1996}. The enthalpy curves (Fig. 2) show the relative thermodynamic stabilities of the predicted structures. The hexagonal $P6_122$ structure [Fig. 1(a), see Ref.~\onlinecite{Ouyang2004}] is most stable at ambient pressure for all of these compounds. Under pressure, however, we find a group of new oxynitride phases.

At 30 GPa, the most stable structures for LaSiO$_2$N, LaGeO$_2$N, and BiSiO$_2$N are predicted to be tetrahedrally coordinated trigonal polar  phases [Fig. 2(d)-(f)]. These structures are piezoelectric, but cannot be ferroelectric by symmetry. Nonpolar, centrosymmetric, perovskite phases are favorable above $\sim$18 GPa and $\sim$7 GPa for YSiO$_2$N ($I4/mcm$) and BiSiO$_2$N ($Cmcm$), respectively.  YGeO$_2$N and YHfO$_2$N stabilize into non-perovskite structures under high pressures. The structural parameters for these structures can be found in the supplemental material~\cite{supp}. We have computed the phonon dispersion for all of these structures, and find that these compounds are dynamically stable [Fig. S1].  

We find that the $P4mm$ structure of BiSiO$_2$N is very close in enthalpy to the most stable structure under pressure. Perovskite phase stability is often governed by the radius ratio of the A and B cations~\cite{Goldschmidt}. Indeed we find that the larger value of r$_A$/r$_B$, the more stable the $P4mm$ phase, where r$_A$ and r$_B$ are the ionic radii~\cite{Shannon1969} of atoms at A and B site, respectively [Fig. 3(a)]. Although the $P4mm$ BiSiO$_2$N is relatively more stable in enthalpy than other compositions, the imaginary phonon frequencies show that it is dynamically unstable. The second stable $P4mm$ structure is in LaSiO$_2$N, and it is only 90 meV / f.u. higher in enthalpy than the $P3_121$ structure at 30 GPa.

\begin{figure}
    \includegraphics[width=8.5cm]{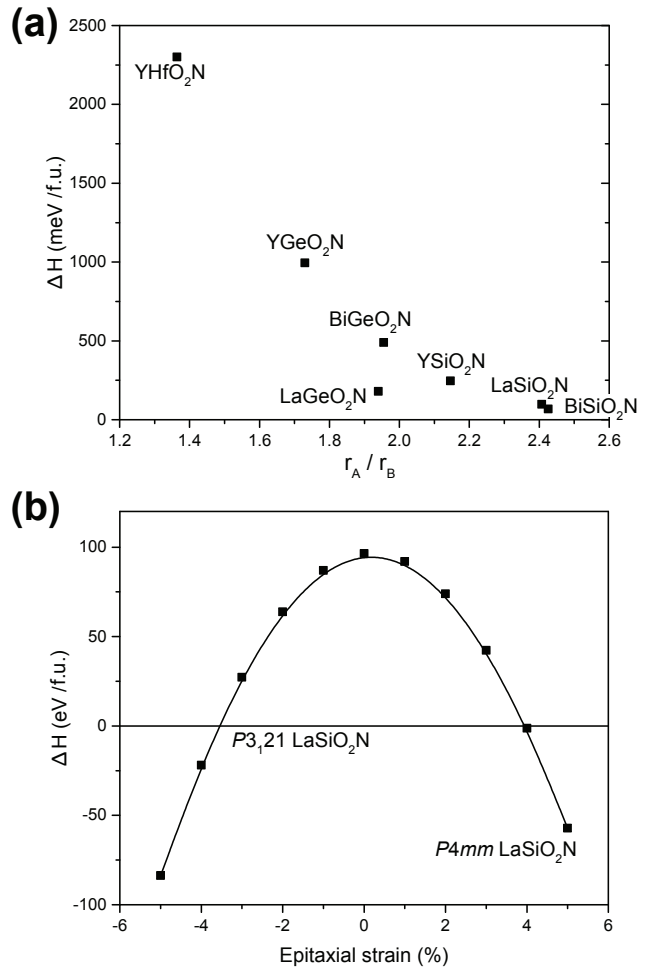}
    \caption{\label{fig3} (a) The calculated enthalpies of the $P4mm$ structure with respect to the most stable structure of the various oxynitrides at 30 GPa. (b) The calculated enthalpies of the $P4mm$ structure of LaSiO$_2$N as a function of epitaxial strain with respect to the $P3_121$ structure at 30 GPa. } 
\end{figure}

The $P4mm$ ferroelectric phase of YSiO$_2$N was calculated to be metastable with the local density approximation~\cite{Hohenberg1964,Ceperley1980} and Troullier-Martins pseudopotentials~\cite{Troullier1991} in a previous study~\cite{Caracas2007}. However, we find that the $P4mm$ structure of YSiO$_2$N is dynamically unstable with the PBEsol functional~\cite{Perdew2008} in the current study, although we were able to reproduce the earlier results using the same pseudopotentials. Nevertheless, synthesis of the $P4mm$ structure of YSiO$_2$N was reported under high pressure and temperature conditions~\cite{Vadapoo2017}. How can one explain its apparent stability? We found that the $I4cm$ structure of YSiO$_2$N [Fig. 1(d)] is slightly lower in energy than the $P4mm$ structure [Fig. 2(a)], and is dynamically stable [Fig. S4]. The $I4cm$ YSiO$_2$N shares nearly the same XRD patterns with the the $P4mm$ phase [Fig. S5]. There is a possibility that the $I4cm$ structure of YSiO$_2$N was actually synthesized rather than the $P4mm$ structure. However, our enthalpy calculations suggest that the $I4cm$ YSiO$_2$N is unstable under high pressures. At the pressures above 10 GPa, the $I4cm$ structure transforms into the most stable $I4/mcm$ structure after geometry optimization [Fig. 2(a)], suggesting that there is no energy barrier between these two phases. Thus $I4cm$ might form on decompression of I4mcm YSiO$_2$N as pressure is decreased. So again, how can we explain the report of polar tetragonal perovskite YSiO$_2$N  observed \textit{in situ} at high pressures \cite{Vadapoo2017}? 
YN was used as one of the precursors in the experimental synthesis. YN adopts the rocksalt-type face-centered cubic structure, and the lattice of YSiO$_2$N can share N atoms along the [110] directions on surface of YN [Fig. S6]. The lattice mismatch between YN and YSiO$_2$N induces around 2\% epitaxial strain, which could stabilize the $P4mm$ structure of YSiO$_2$N. Previous studies have shown that epitaxial strain can stabilize ferroelectric structures or trigger spontaneous polarization in perovskite oxides~\cite{Haeni2004,Choi2004,Liu2016c,Xu2020}. To investigate the effect of applied epitaxial strain on the stability of perovskite oxynitrides, we fix the lattice parameters of $a$ and $b$ with -5\% to 5\% strains, and relax $c$ axis and atomic positions. We next calculate the phonon dispersion of these strained structures, which reveal that the extensional strain hardens the phonon modes. Thus we propose that the successful synthesis of the $P4mm$ structure of YSiO$_2$N can be ascribed to epitaxial strain on YN.

Nonhydrostatic pressure has previously been found to promote formation of metastable materials. For example, a large uniaxial stress was found to promote the synthesis of the cubane-derived nanothreads in the diamond anvil cell~\cite{cubane}. Since nonhydrostatic pressure or epitaxial strain might promote formation of ferroelectric oxynitrides, we studied our candidates polar oxynitrides under epitaxial strain (non-hydrostatic pressure) as well as hydrostatic pressure.

\begin{figure}
    \includegraphics[width=8.5cm]{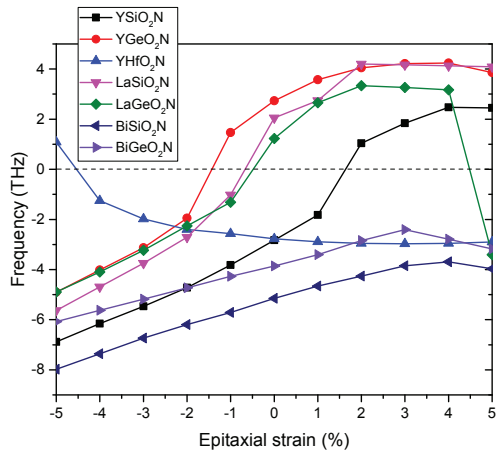}
    \caption{\label{fig4} Soft phonon modes for the $P4mm$ structures of the oxynitrides as functions of the epitaxial strain at ambient pressure. } 
\end{figure}

The softest phonon modes for the $P4mm$ structures of the oxynitrides as a function of epitaxial strain indicates which compounds and conditions are at least metastable [Fig. 4]. 
We find that the $P4mm$ perovskite structures of YGeO$_2$N, LaSiO$_2$N, and LaGeO$_2$N are dynamically stable at zero applied stress [Fig. S7], and the phonon modes harden with the increasing stress. The computed polarizations of 2.21 C/m$^2$ (YGeO$_2$N), 1.27 C/m$^2$ (LaSiO$_2$N), and 0.97 C/m$^2$ (LaGeO$_2$N) are much larger than that of BaTiO$_3$, and the values of the polarization are among the highest ever reported so far in the literature.
The dynamical stability of these ferroelectric phases makes it likely they can be synthesized at ambient pressure using layer-by-layer growth methods such as pulzed laser deposition (PLD)~\cite{smith1965vacuum}, molecular beam epitaxy (MBE)~\cite{cho1975molecular} or chemical vapor deposition (CVD) etc., which have been used to synthesize many metastable materials, including diamond and carbon nanotubes~\cite{Meng2008a,Cassell1999}. For LaGeO$_2$N, the structure collapsed when the epitaxial strain reaches 5\%, resulting in the imaginary phonon frequencies again.  The $P4mm$ phases of BiSiO$_2$N and BiGeO$_2$N remain dynamically unstable at the epitaxial strain of -5\% to 5\%. A trend of the modes with the strain for YHfO$_2$N is opposite to the others, the blue curve in Fig. 4.  The compressive strain by -5\% stabilizes the $P4mm$ structure of YHfO$_2$N. 

We performed first-principles molecular dynamics (FPMD) for  $4\sqrt{2}\times 4\sqrt{2} \times 4$  (160 atoms) supercells for temperatures from 300-600 K starting from the  $I4cm$ structure, and found it stable up to 600 K under compressive epitaxial strain and to 400 K under no or extensional strain, so if one could produce tetragonal YSiO$_2$N by layer by layer epitaxial growth, it could be stable and useful if not heated too much.

However, we find other compositions which may be more stable and promising for further study. 
From the enthalpy perspective, $P4mm$ LaSiO$_2$N is promising for synthesis under epitaxial strain at 30 GPa [Fig. 3(b)].
The dynamic stabilities of the $P4mm$ LaSiO$_2$N were examined by calculating the phonon spectra, and no imaginary phonon frequencies were found in the whole Brillouin zone under epitaxial strain of -5\%, -4\%, and 4\% at 30 GPa [Fig. S3]. 
The stability of the strained $P4mm$ LaSiO$_2$N is encouraging for future synthesis efforts. Besides the $P4mm$ LaSiO$_2$N, we found the ferroelectric $Cmc2_1$ structures of BiSiO$_2$N and BiGeO$_2$N are also metastable at ambient pressure [Fig. S8], which are also very promisig for further experimental studies.

In summary, we have combined automatic structure searching methods with first-principles calculations to investigate the phase stability of oxynitrides. We predict that LaSiO$_2$N, LaGeO$_2$N, and BiSiO$_2$N are thermodynamically stable in the polar trigonal perovskite structure at 30 GPa. In addition, our study reveals that the dynamical stability of the ferroelectric phase of oxynitrides is sensitive to the epitaxial strain. The energy calculations suggest that it is promising to synthesize ferroelectric oxynitrides at ambient pressure using nonequilibrium synthesis methods or under high pressure and epitaxial strain. This work also provides a vision for searching synthesizable ferroelectric oxynitrides by controlling the size ratio of different elements in A/B-site of perovskite structures.

This work is supported by U. S. Office of Naval Research Grants No. N00014-17-1-2768 and N00014-20-1-2699, and the Carnegie Institution for Science. Computations were supported by DOD HPC, Carnegie computational resources, and REC gratefully acknowledges the Gauss Centre for Supercomputing e.V. (www.gausscentre.eu) for funding this project by providing computing time on the GCS Supercomputer SuperMUC-NG at Leibniz Supercomputing Centre (LRZ, www.lrz.de).

\bibliography{Oxyref}

\setcounter{figure}{0}
\onecolumngrid
\newpage

\renewcommand{\thetable}{S\arabic{table}}  
\renewcommand{\thefigure}{S\arabic{figure}}

\section*{Supplemental Material for 
``Stable Polar Oxynitrides through Epitaxial Strain''}

\begin{figure}[h]
    \includegraphics[width=12cm]{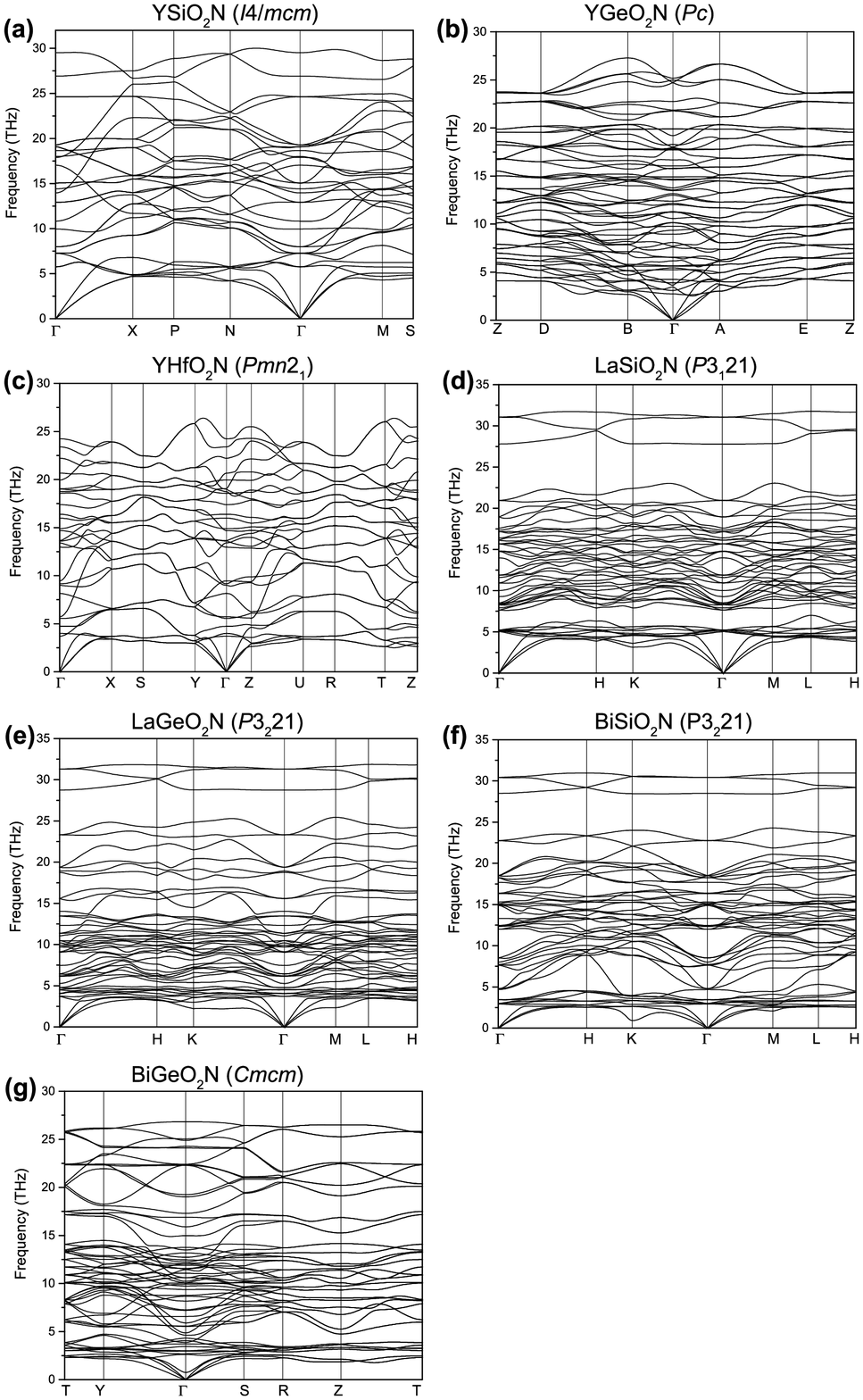}
    \caption{Calculated phonon dispersions of various predicted structures at 30 GPa. }
\end{figure}

\clearpage

\newpage

\begin{figure}
    \includegraphics[width=15cm]{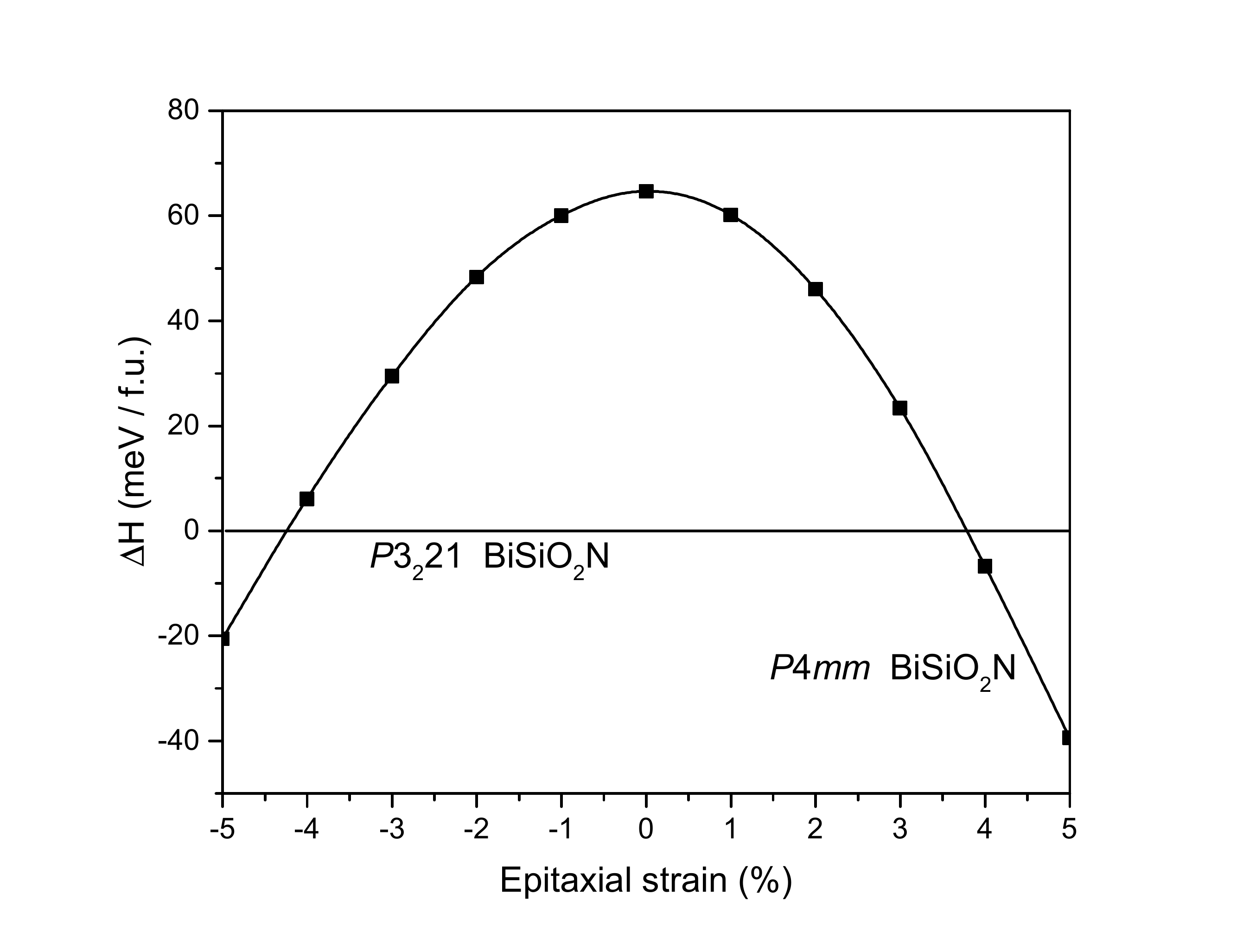}
    \caption{The calculated enthalpies of the $P4mm$ structure of BiSiO$_2$N as a function of epitaxial strain with respect to the $P3_221$ structure at 30 GPa.}
\end{figure}

\clearpage
\newpage

\begin{figure}
\includegraphics[width=15cm]{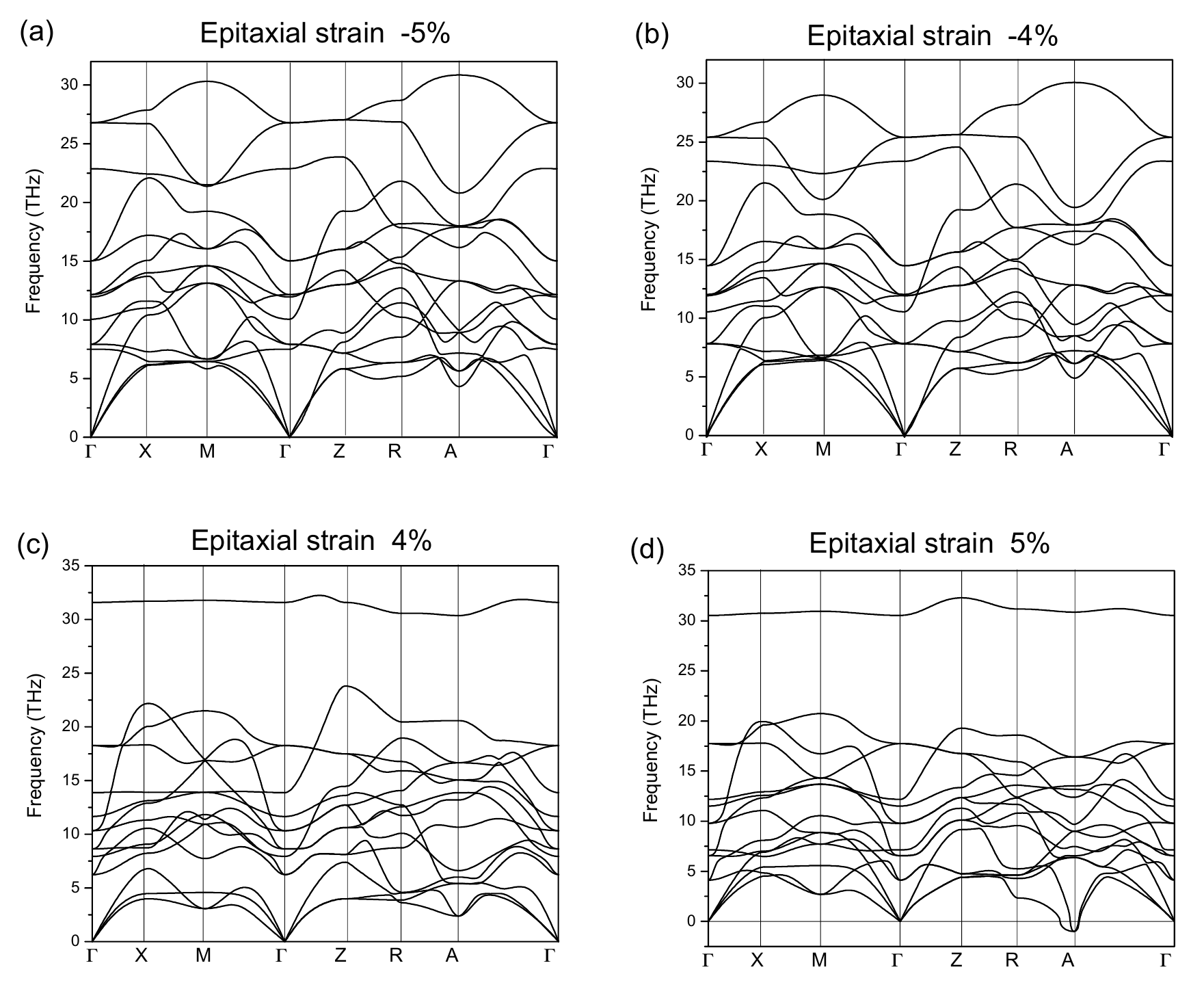}
\caption{Calculated phonon dispersions of the $P4mm$ LaSiO$_2$N under the epitaxial strain of -5\%, -4\%, 4\%, and 5\% at 30 GPa. } 
\end{figure} 
\clearpage
\newpage

\begin{figure}
    \includegraphics[width=15cm]{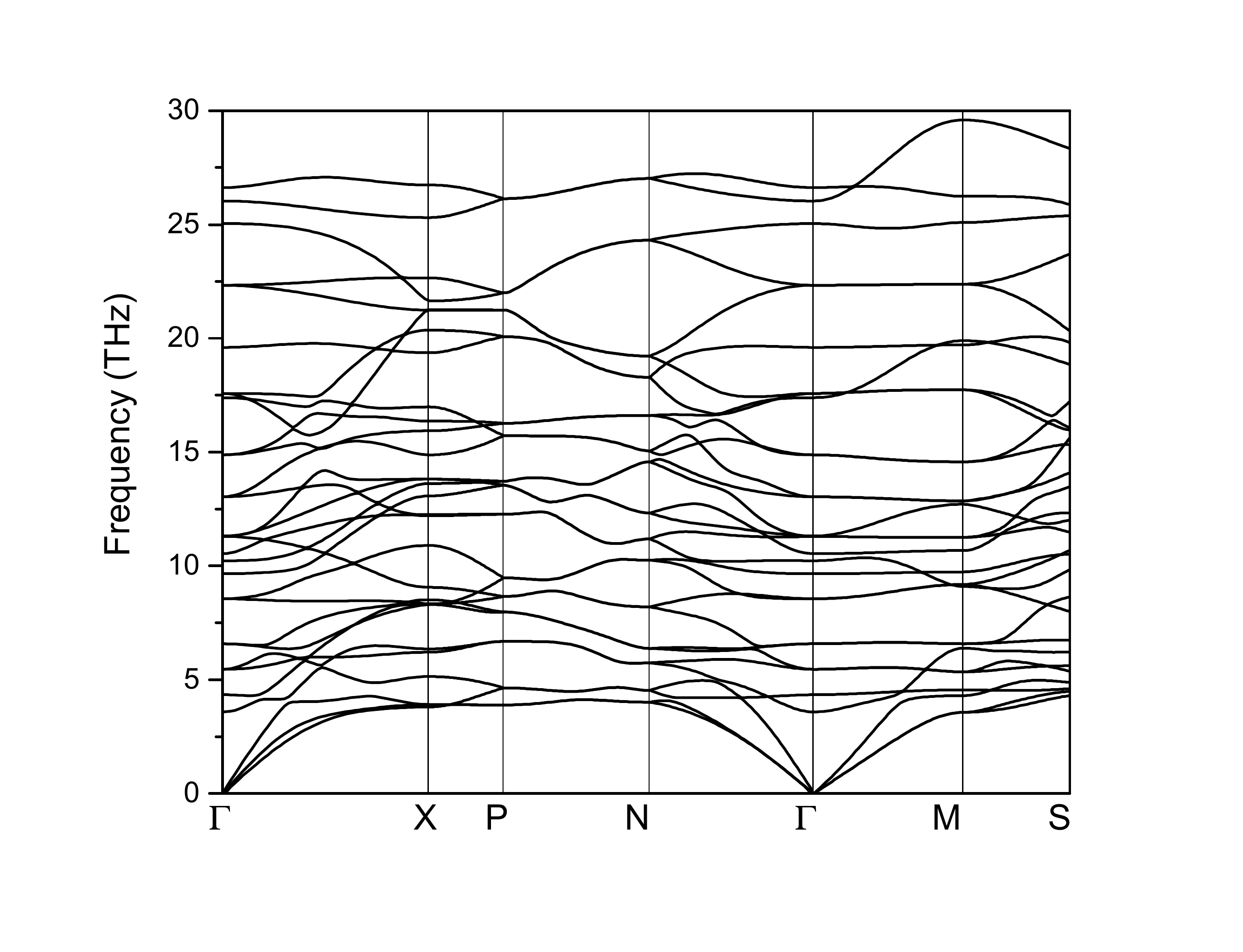}
    \caption{Calculated phonon dispersions of the $I4cm$ YSiO$_2$N at 0 GPa.}
\end{figure}
\clearpage
\newpage

\begin{figure}
    \includegraphics[width=15cm]{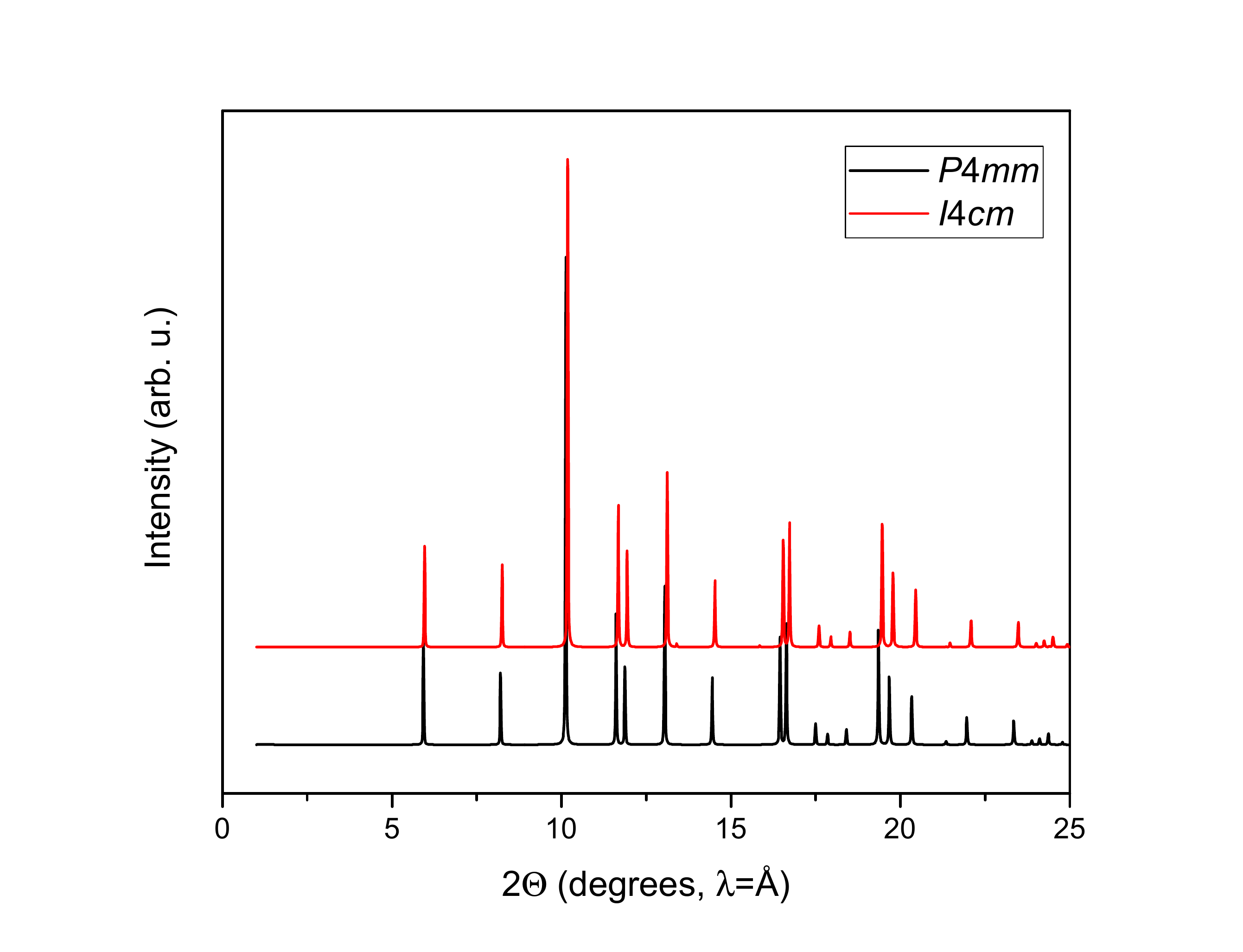}
    \caption{Calculated XRD patterns of the $P4mm$ and the $I4cm$ structures of YSiO$_2$N at 0 GPa.}
\end{figure}
\clearpage
\newpage

\begin{figure}
    \includegraphics[width=15cm]{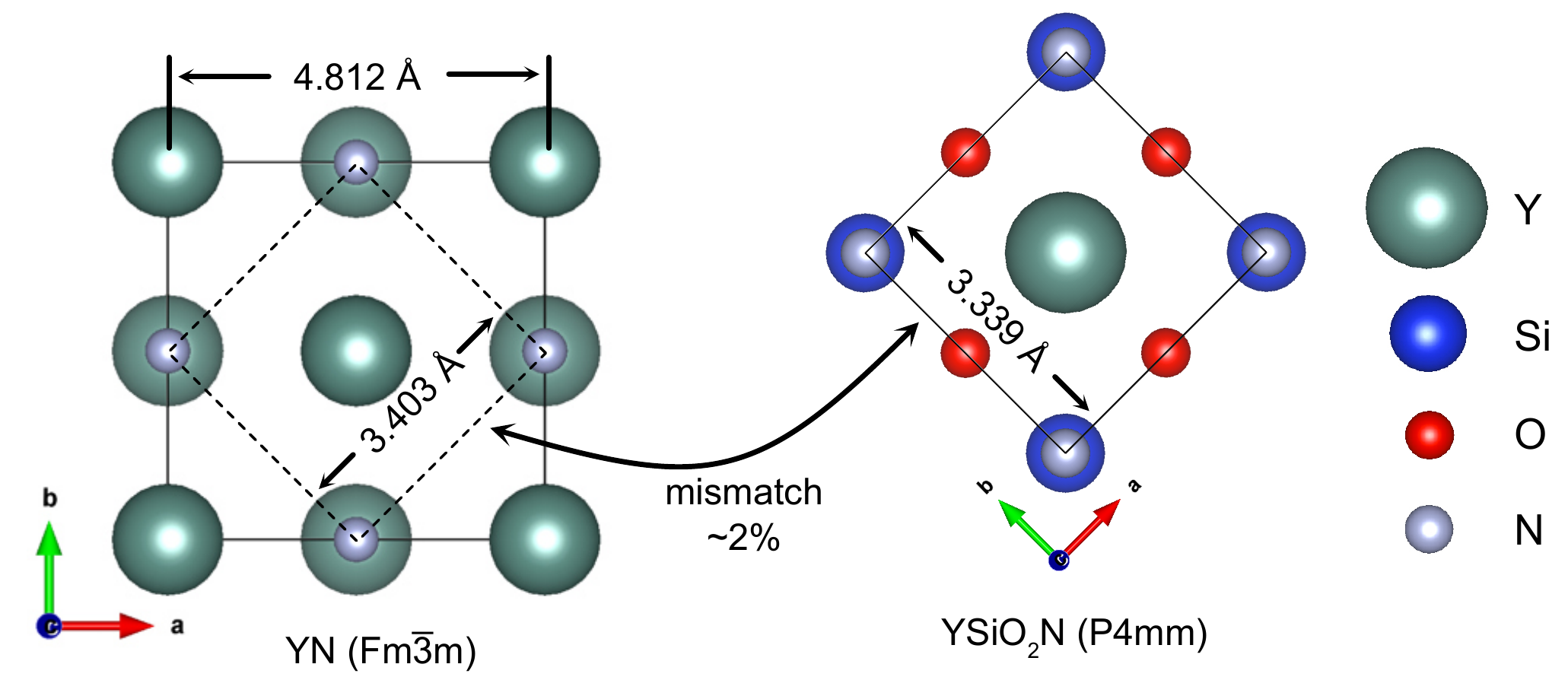}
    \caption{The top view of the crystal structures of YN (left) and YSiO$_2$N (right). The  lattice  mismatch  between  YN  and  YSiO$_2$N induces  around  2\%  epitaxial  strain.} 
    \end{figure} 
    \clearpage
    \newpage

\begin{figure}
    \includegraphics[width=8cm]{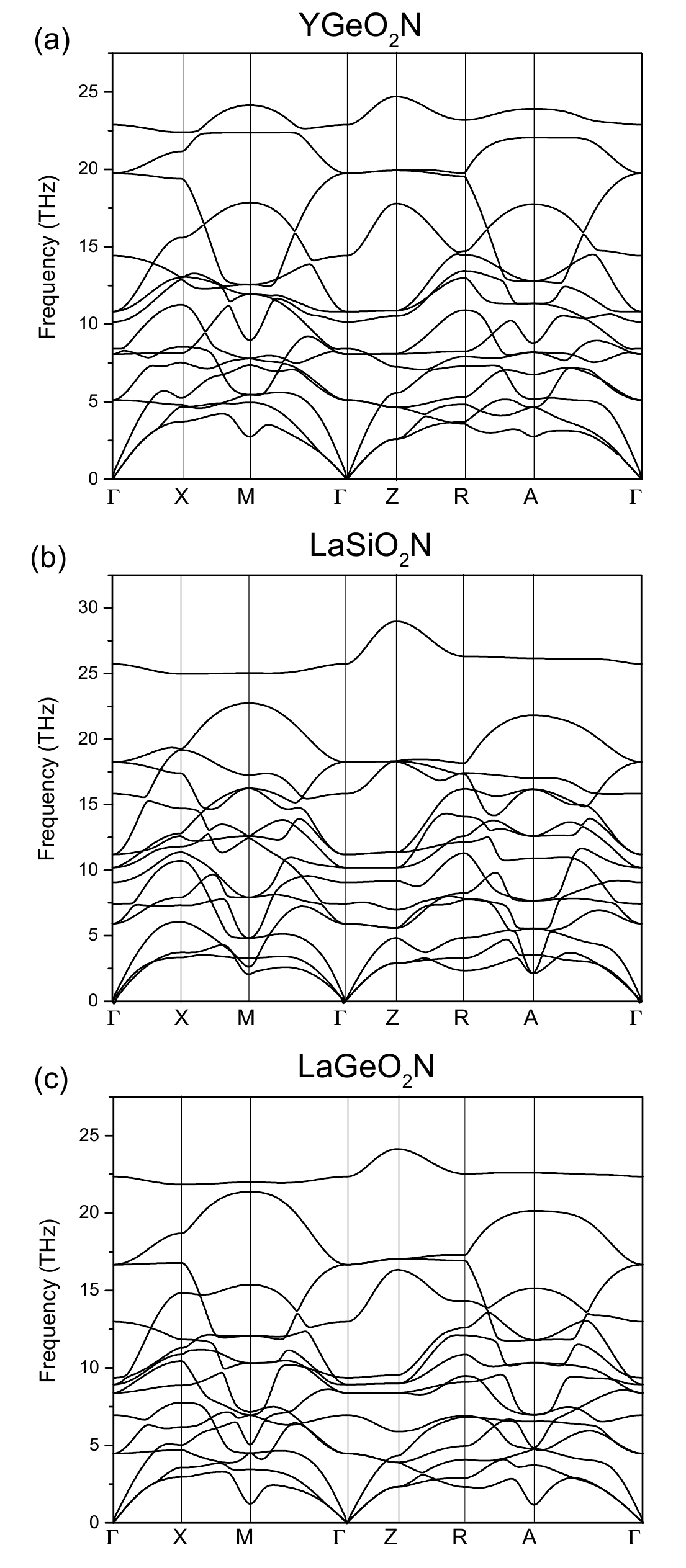}
    \caption{Calculated phonon dispersions of the $P4mm$ structure of YGeO$_2$N (a), LaSiO$_2$N (b), and LaGeO$_2$N (c) at 0 GPa.} 
    \end{figure} 
    \clearpage
    \newpage

\begin{figure}
    \includegraphics[width=8cm]{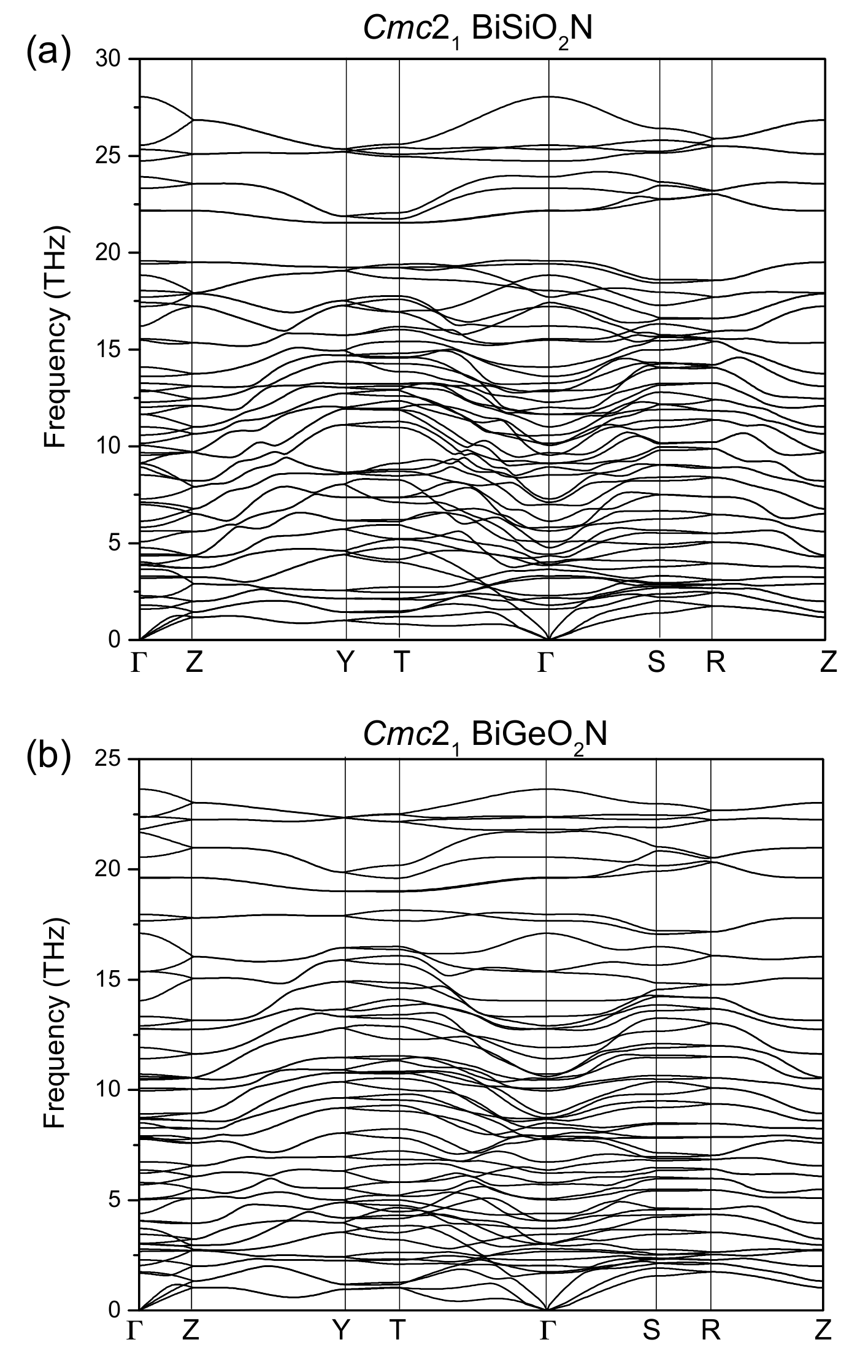}
    \caption{Calculated phonon dispersions of the $Cmc2_1$ structure of BiSiO$_2$N (a), and BiGeO$_2$N (b) at 0 GPa.} 
    \end{figure} 
    \clearpage
    \newpage    

\begin{longtable}{ccccccc}
    \caption{ Calculated structural parameters for various structures of oxynitrides.}\\
\hline
\hline
     & Space Gruop& Lattice parameters (\AA, $^\circ$) &\multicolumn{4}{c}{Atomic coordinates (fractional)} \\ \hline
     & & & Atom & X & Y & Z \\ \hline
    YSiO$_2$N & $I4/mcm$ & a = 4.897 & Y($4b$) & 0 & \textonehalf & \textonequarter \\ 
    30 GPa & & b = 7.306 & Si($4c$) & 0 & 0 & 0 \\ 
    & & & O($8h$) & 0.6979 & 0.1979 & 0 \\ 
    & & & N($4a$) & 0 & 0 & \textonequarter \\ \hline

    YSiO$_2$N & $I4cm$ & a = 4.756 & Y($4b$) & \textonehalf & 0 & 0.1548\\
    0 GPa & & c = 9.304 & Si($4a$) & 0 & 0 & 0.3967 \\
    & & & O ($8c$) & 0.7743 & 0.2743 & 0.3364 \\
    & & & N ($4a$) &  0 & 0 & 0.0798 \\ \hline

    YGeO$_2$N & $Pc$ & a = 5.875 & Y1($2a$) & 0.7816 & 0.9997 & 0.0767 \\
    30 GPa & & b = 2.983 & Y2($2a$) & 0.2183 &  0.4994 & 0.4234 \\
    & & c = 11.493 & Ge1($2a$) & 0.3389 &  0.5021 &  0.6767 \\
    & & $\beta$ = 99.08 & Ge2($2a$) & 0.6609 &  0.0022 &  0.8234 \\
    & & & O1($2a$) & 0.0483 &0.4997 &  0.0998 \\
    & & & O2($2a$) & 0.9515& 0.0007 & 0.9001 \\
    & & & O3($2a$) & 0.6155& 0.5007 & 0.9383\\
    & & & O4($2a$) &  0.3843&  0.0005 &  0.0619 \\
    & & & N1($2a$) & 0.3266& 0.9984 &0.2848 \\
    & & & N2($2a$) & 0.6732&  0.4985&  0.2154 \\ \hline

    YHfO$_2$N & $Pmn2_1$ & a = 3.135 & Y($2a$) & 0 & 0.3608 & 0.1104 \\
    30 GPa & & b = 6.537 & Hf($2a$) & 0 & 0.1004 & 0.6131  \\
    & & c = 5.158 & O1($2a$) & 0 & 0.8300 & 0.8404 \\
    & & & O2($2a$) & 0 & 0.4230 & 0.5680 \\
    & & & N($2a$) & 0 & 0.8390 & 0.3685\\ \hline
    LaSiO$_2$N & $P3_121$ & a = 5.124 & La($3a$) & 0.3303 &  0 & \sfrac{1}{3} \\
    30 GPa & & c = 6.266 & Si($3b$) & 0.3160 &  0 & \sfrac{5}{6} \\ 
    & & & O($6c$) & 0.6672 &  0.1723 & 0.9884 \\
    & & & N($3a$) & 0.8221 &  0 & \sfrac{1}{3} \\ \hline

    LaGeO$_2$N & $P3_221$ & a = 5.296 & La($3a$) & 0.6739 &  0 & \sfrac{2}{3} \\
    30 GPa & & c = 6.499 & Ge($3b$) & 0.6785 &  0 & \sfrac{1}{6} \\ 
    & & & O($6c$) & 0.4963 &  0.6875 & 0.3581 \\
    & & & N($3a$) & 0.1998 &  0 & \sfrac{2}{3} \\ \hline

    BiSiO$_2$N & $P3_221$ & a = 5.296 & Bi($3a$) & 0.6695 &  0 & \sfrac{2}{3} \\
    30 GPa & & c = 6.499 & Si($3b$) & 0.6759 &  0 & \sfrac{1}{6} \\ 
    & & & O($6c$) &0.5029 & 0.1919 & 0.3115 \\
    & & & N($3a$) & 0.2014 &  0 & \sfrac{2}{3} \\ \hline

    BiSiO$_2$N & $Cmc2_1$ & a = 7.062 & Bi1($3a$) & 0 &  0.7564 & 0.8660 \\
    0 GPa & & b = 6.519 & Bi2($4a$) & 0 &  0.2608 & 0.8484 \\ 
     & & c = 9.893 & Si($8b$) & 0.7451 &  0.0121 & 0.6034 \\ 
     & & & O1($8b$) & 0.7236 &  0.2641 & 0.6668 \\ 
     & & & O2($4a$) & 0 &  0.4860 & 0.6739 \\ 
     & & & O3($4a$) & 0 &  -0.0478 & 0.1001 \\ 
     & & & N($8b$) & 0.2085 &  0.4962 & -0.0686 \\  \hline

    BiGeO$_2$N & $Cmcm$ & a = 7.369 & Bi1($4c$) & 0 &  0.9768 & \sfrac{1}{4} \\
    30 GPa& & b = 7.386 & Bi2($4c$) & 0 &  0.5074 & \sfrac{1}{4} \\ 
    & & c = 7.711  & Ge($8d$) & \sfrac{1}{4} & \sfrac{1}{4} & 0 \\
    & & & O1($8f$) & 0 &  0.1955 & 0.0172 \\ 
    & & & O2($8e$) & 0.3052 &  0 & 0 \\ 
    & & & N($8g$) & 0.7138& 0.2466& \sfrac{1}{4} \\ 
    \hline
    \hline
    \end{longtable}

\end{document}